\documentstyle[12pt]{article}
\font\twlmsy=msbm10 at 12pt
\font\sevenmsy=msbm8
\font\fivemsy=msbm6
\newfam\Bbbfam
\textfont\Bbbfam=\twlmsy
\scriptfont\Bbbfam=\sevenmsy
\scriptscriptfont\Bbbfam=\fivemsy

\newcommand{\rf}[1]{(\ref{#1})}

\newcommand{\beq}{\begin{equation}}
\newcommand{\eeq}{\end{equation}}
\newcommand{\bea}{\begin{eqnarray}}
\newcommand{\eea}{\end{eqnarray}}
\newcommand{\beas}{\begin{eqnarray*}}
\newcommand{\eeas}{\end{eqnarray*}}
\newcommand{\beqs}{\begin{displaymath}}
\newcommand{\eeqs}{\end{displaymath}}

\newcommand{\ul}{\underline}

\newcommand{\cT}{{\cal T} }
\newcommand{\cZ}{{\cal Z} }

\newcommand{\ben}{\begin{equation}}
\newcommand{\een}{\end{equation}}

\newcommand{\bdm}{\begin{displaymath}}
\newcommand{\edm}{\end{displaymath}}

\newcommand{\pa}{\partial}

\begin{document}
\topmargin 0pt
\oddsidemargin 5mm
\headheight 0pt
\topskip 0mm

\addtolength{\baselineskip}{0.3\baselineskip}

\pagestyle{empty}

\hfill RH-18-97

\vspace{2 truecm}

\begin{center}

{\Large \bf A polymer gas on a random surface}

\vspace{1 truecm}


\vspace{1.2 truecm}

{\em Bergfinnur Durhuus\footnote{e-mail: durhuus@math.ku.dk}}

\bigskip

Matematisk Institut, University of Copenhagen\\
Universitetsparken 5, 2100 Copenhagen\\
Denmark

\vspace{1.2 truecm}

{\em Thordur Jonsson\footnote{e-mail: thjons@raunvis.hi.is} }

\bigskip

Raunvisindastofnun Haskolans, University of Iceland \\
Dunhaga 3, 107 Reykjavik \\
Iceland

\vspace{.2 truecm}

\end{center}

\vfill

\noindent
{\bf Abstract.} Using the observation that configurations of 
$N$ polymers with hard core interactions on a closed random surface 
correspond to random 
surfaces with $N$ boundary components we calculate the free 
energy of a gas of polymers interacting with fully quantized
two-dimensional gravity.
 We derive the equation  of state for the polymer gas 
and find that all the virial coefficients beyond the second one 
vanish identically.  

\vfill

\newpage
\pagestyle{plain}

It is well known that some statistical mechanical models are easier to solve on 
a random triangulated 
surface than on a fixed lattice.  Well known examples are the Ising model
\cite{kazakov} and self avoiding random walks \cite{duplantier1}.  In the
latter reference a closed form for 
the partition function of $n$ mutually self avoiding 
walks with common endpoints (watermelon diagrams) 
was obtained and the critical exponents determined; see \cite{duplantier2} for
a review.

The reason for the simplicity of self avoiding random walks on a random 
surface is that we can keep the walks fixed and sum over the surfaces in
a relatively straightforward fashion.  One can for example 
regard the links which make up the walks as boundary 
components where the links have been identified pairwise.  In fact one can also
identify the boundary links in such a way that one obtains a branched polymer on
the surface.  The statistical sum over surfaces and walks therefore becomes a 
sum over surfaces with fluctuating boundaries.  

In \cite{ambjorn1}  
closed expressions were derived for all the multiloop functions for 
random surfaces made up of even sided polygons. 
The scaling limit of the loop functions in these models 
was also calculated and it was shown that the limit is
the same when one drops the assumption that the polygons are even sided.
The authors of \cite{ambjorn1} used 
matrix model techniques.   Identical results 
can be obtained working directly with triangulations \cite{book}.
Here we use the explicit results of \cite{ambjorn1} 
to study the thermodynamical properties of
a gas of polymers interacting with 2d quantum gravity.  This might be
regarded as a two-dimensional 
toy model for primordial cosmic strings in thermal equilibrium with 
the gravitational field.

Let us consider an ensemble of triangulated surfaces with the topology of $S^2$ 
with $b$ boundary components.  By a triangulation we here mean a collection
of polygons whose sides are identified pairwise and degenerate boundaries may 
appear (unrestricted triangulations in the terminology of \cite{book}).
We assume that each boundary component consists of an even number of links and 
carries one marked link.  Let us arrange the links cyclically along each
boundary component: 
$i_1,i_2,\ldots i_{2l}$, so that $i_1$ is the marked link.  If we    
identify the link $i_1$ with $i_{2l}$, $i_2$ with $i_{2l-1}$ etc.\ we 
obtain a 
self-avoiding walk on a surface with one less boundary component.
One of the enpoints of the walk is marked corresponding to the marking
of the original boundary component.  
Doing this for all the boundary components we end up with a collection of $b$ 
mutually self-avoiding walks.
We shall call the self-avoiding walks linear polymers or just polymers
from now on.  
 Given an orientation of the surface we see that there are precisely two 
distinct markings of a boundary component 
that give rise to the same unmarked
linear polymer.
Similarly one can obtain a branched polymer on the surface by identifying the 
boundary links in a more general fashion.  This can be done in precisely
\beq
C_l={(2l)!\over l!(l+1)!}
\eeq
different ways, see e.g.\ \cite{book}.  
Here we shall focus on the case of linear 
polymers. The general case can be treated by the same 
methods and we give the result below.

It follows from the above considerations that we can write the partition 
function of a gas consisting of $b$ 
polymers of lengths $l_1,l_2,\ldots ,l_b$ with a hard core interaction 
on a random surface as
\beq
\cZ _b(l_1,\ldots ,l_b)=2^{-b}w(2l_1,\ldots ,2l_b),\label{2}
\eeq
where $w$ is the $b$-loop function of quantum gravity (discretized 
Hartle--Hawking wave functional).  The loop function is given by
\beq
w(l_1,l_2,\ldots ,l_b)=\sum _{T\in\cT (l_1,\ldots l_b)}
e^{-S_T}
\eeq
where $\cT (l_1,\ldots ,l_b)$ is the collection of all triangulations in the 
class under study which have 
$b$ boundary components of lengths $l_1,\ldots ,l_b$ 
and $S_T$ is the action of the triangulation $T$.

We briefly recall the derivation of the formula for the loop function.  
For simplicity we consider triangulations which are made up of 
even sided polygons and take the action
to be of the form
\beq
S_T=-\sum_{j=1}^\infty k_j\log g_{2j},
\eeq
where $k_j$ is the number of $2j$-gons in $T$.  Only a finite number of the 
coupling constants $g_{2j}$ 
are nonzero and they are assumed to be nonnegative since we are interested in 
the pure gravity case but in principle one can study the multicritical models
by the same method.  Denoting the coupling constants collectively by
$\ul g$ we can write the Laplace--transform of the $b$-loop function as
\beq
\hat{w}(\ul{g};z_1,\ldots ,z_b)=\prod_{i=2}^b{d\over dV(z_i)}\; \hat{w}(\ul{g}
;z_1),
\eeq
where
\beq
\hat{w}(\ul{g};z)=\sum_{l=1}^\infty z^{-2l-1}w(\ul{g};2l)
\eeq
 is the Laplace--transform of the one-loop function and the ``loop insertion 
operator'' is defined by
\beq
{d\over dV(z)}=\sum_{j=1}^\infty {d\over dg_{2j}},
\eeq
see \cite{ambjorn1,book}.  One should regard the variables $z_i$ 
as the cosmological constants for boundaries.  After some calculation one 
obtains for $b\geq 3$ the formula
\beq
\hat{w}(\ul{g};z_1,\ldots ,z_b)=\left({2\over 2\tilde{M}_1(c^2)}{d\over dc^2}
\right)^{b-3}{1\over 2c^2\tilde{M}_1(c^2)}\prod_{k=1}^b
{c^2\over (z_k^2-c^2)^{3/2}},
\eeq
where $c=c(\ul g)$ is the critical value of the $z_i$-variables,
\beq
\tilde{M}_1(c^2)=\oint {d\omega\over 2\pi i}{\omega V'(\omega )\over
(\omega ^2-c^2)^{3/2}}
\eeq
and
\beq
V(\omega )=\sum _{i=1}^\infty g_{2j}\omega ^{2j}.
\eeq
Writing $g_{2j}=gt_{2j}$ where $g>0$ and the $t_{2j}$ are kept fixed we 
have a critical value $g_0$ of $g$ where the loop functions become singular.

The scaling limit is calculated by introducing a lattice length $a$ which
tends to zero and letting
$g\to g_0$ and $c\to c_0\equiv c(g_0)$ such that
\bea
\log g-\log g_0&=&a^2\Lambda \\
\log z_i- \log c_0&=&aZ_i,
\eea
where $\Lambda$ and $Z_i$ are renormalized dimensionfull 
cosmological constants which are kept fixed in the scaling limit.
The continuum loop functions are given by
\bea
\widehat{W}(\Lambda ;Z_1,\ldots ,Z_b) & \equiv &\lim_{a\to 0}
a^{-5+7b/2}\hat{w}(\ul{g};z_1,\ldots ,z_b) \nonumber \\
  &=& C\left({d\over d\Lambda }\right)^{b-3}{1\over \sqrt{\Lambda}}\prod
_{i=1}^b{1\over (Z_i+\sqrt{\Lambda})^{3/2}},\label{final}
\eea
for $b\geq 3$ and $C$ is a numerical constant.  There are also explicit 
formulas for $b=1$ and $2$ but they are not of interest to us here.

It is now straightforward to take the inverse 
Laplace--transform in the variables $Z_i$ and 
$\Lambda$ and this yields the 
continuum $b$-loop function at fixed volume $V$ and loop 
lengths $L_1,\ldots 
,L_b$:
\beq
W(V;L_1,\ldots ,L_b)=CV^{b-7/2}\sqrt{\prod_{i=1}^bL_i}\;
\exp\left( -{1\over 4V}
\left(\sum_{i=1}^bL_i\right)^2\right).\label{14}
\eeq

Let now all the polymers have the same continuum length $L$.  Then the 
partition function becomes
\beq
\cZ _b(V,L)=C{V^{b-7/2}\over b!}L^{b/2}e^{-b^2L^2/4V},
\eeq
where we have inserted the 
statistics factor $(b!)^{-1}$  in order to ensure 
``correct Boltzmann counting'' of polymers.  Putting
\beq
\rho ={b\over V}
\eeq
and taking the limit $b\to\infty$ and $V\to\infty$ 
with the polymer density 
$\rho$ fixed we obtain the free energy per unit volume  
\beq\label{free}
f(\rho ,L)=\rho\log\rho -{\rho\over 2}\log L+{\rho^2L^2\over 4}.
\eeq
The first term on the right hand side of 
\rf{free} corresponds to an ideal gas.

The pressure is given by
\bea
p &=& {\pa\over\pa\rho ^{-1}}\left( \rho ^{-1}f(\rho ,L)\right)
\nonumber \\
         &=&\rho +{\rho ^2L^2\over 4}.\label{es}
\eea
The equation \rf{es} is the exact equation of state for the polymer gas and 
we see in particular that the hard core repulsion of polymers is reflected 
in a single quadratic term in the virial expansion of the pressure.  This
is much simpler than the corresponding formula for the pressure for polymers 
in flat space where in general all the virial coefficients are non-vanishing 
\cite{pol1,pol2}.
It is clear that this gas has no phase transition as expected for a gas with 
purely repulsive interaction.

In the case of general branched polymers we obtain an extra entropy
factor
\beq
\prod _{i=1}^{b}C_{l_i}
\eeq
on the right hand side of \rf{2}.  Using
\beq
C_l\sim 4^l l^{-3/2}
\eeq
for large $l$ we find that the continuum partition function at fixed
volume is in this case given by
\beq
Z_b'(V,L)=C{V^{b-7/2}\over b!}L^{-b}e^{-b^2L^2/4V},
\eeq
which yields the free energy
\beq
f'(\rho ,L)=\rho\log\rho +\rho\log L+{\rho^2L^2\over 4}
\eeq
and the equation of state is unchanged.

Using the semi-explicit formulas for the higher genus loop 
functions derived in
\cite{book} one can in principle 
generalize the above calculation to obtain the free 
energy of a polymer gas on a surface of arbitrary fixed genus.  
It is not straightforward to obtain an explicit expression in this case
since there is no known analogue of \rf{14} for higher genus surfaces.
The problem of letting the number of handles go to infinity in the
thermodynamic limit so that one has a finite density of ``wormholes'' can 
presumably also be studied by this method.
 
\bigskip
\noindent
{\bf Acknowledgement.}  This work was partly supported by a NorFa grant.

\end{document}